\documentclass[twocolumn,aps,prl,showpacs,groupedaddress,floatfix,letterpaper,superscriptaddress,reprint]{revtex4}
\usepackage{ae}
\usepackage[utf8]{inputenc}
\usepackage{graphicx}
\usepackage{hyperref}
\usepackage[leqno]{amsmath}
\usepackage{amsfonts}
\usepackage{xspace}
\usepackage[usenames,dvipsnames]{color}
\definecolor{Lila}{rgb}{0.5,0.2,0.5}
\definecolor{Red}{rgb}{0.7,0.0,0.0}
\definecolor{Green}{rgb}{0.0,0.45,0.0}
\newcommand{\BePlus}{$^9$Be$^+$\xspace}

\begin{document}
\title{Versatile control of \BePlus ions using a spectrally tailored UV frequency comb}

\author{A.-G.~Paschke}
\affiliation{Institute of Quantum Optics, Leibniz Universit\"at Hannover, Welfengarten 1, 30167 Hannover, Germany}
\affiliation{Physikalisch-Technische Bundesanstalt, Bundesallee 100, 38116 Braunschweig, Germany}
\author{G.~Zarantonello}
\affiliation{Institute of Quantum Optics, Leibniz Universit\"at Hannover, Welfengarten 1, 30167 Hannover, Germany}
\affiliation{Physikalisch-Technische Bundesanstalt, Bundesallee 100, 38116 Braunschweig, Germany}
\author{H.~Hahn}
\affiliation{Institute of Quantum Optics, Leibniz Universit\"at Hannover, Welfengarten 1, 30167 Hannover, Germany}
\affiliation{Physikalisch-Technische Bundesanstalt, Bundesallee 100, 38116 Braunschweig, Germany}
\author{T.~Lang}
\affiliation{Institute of Quantum Optics, Leibniz Universit\"at Hannover, Welfengarten 1, 30167 Hannover, Germany}
\author{C.~Manzoni}
\affiliation{IFN-CNR, Dipartimento di Fisica, Politecnico di Milano, Piazza L. da Vinci 32, Milano, 20133, Italy}
\author{M.~Marangoni}
\affiliation{IFN-CNR, Dipartimento di Fisica, Politecnico di Milano, Piazza L. da Vinci 32, Milano, 20133, Italy}
\author{G.~Cerullo}
\affiliation{IFN-CNR, Dipartimento di Fisica, Politecnico di Milano, Piazza L. da Vinci 32, Milano, 20133, Italy}
\author{U.~Morgner}
\affiliation{Institute of Quantum Optics, Leibniz Universit\"at Hannover, Welfengarten 1, 30167 Hannover, Germany}
\author{C.~Ospelkaus}
\affiliation{Institute of Quantum Optics, Leibniz Universit\"at Hannover, Welfengarten 1, 30167 Hannover, Germany}
\affiliation{Physikalisch-Technische Bundesanstalt, Bundesallee 100, 38116 Braunschweig, Germany}

\begin{abstract}
We demonstrate quantum control of \BePlus ions directly implemented by an optical frequency comb. Based on numerical simulations of the relevant processes in \BePlus for different magnetic field regimes, we demonstrate a wide applicability when controlling the comb's spectral properties. We introduce a novel technique for the selective and efficient generation of a spectrally tailored narrow-bandwidth optical frequency comb near 313\,nm. We experimentally demonstrate internal state control and internal-motional state coupling of \BePlus ions implemented by stimulated-Raman manipulation using a spectrally optimized optical frequency comb. Our pulsed laser approach is a key enabling step for the implementation of quantum logic and quantum information experiments in Penning traps.

\end{abstract}


\maketitle
\begin{figure}[tb]
	\centering
	\includegraphics[width=0.85\columnwidth]{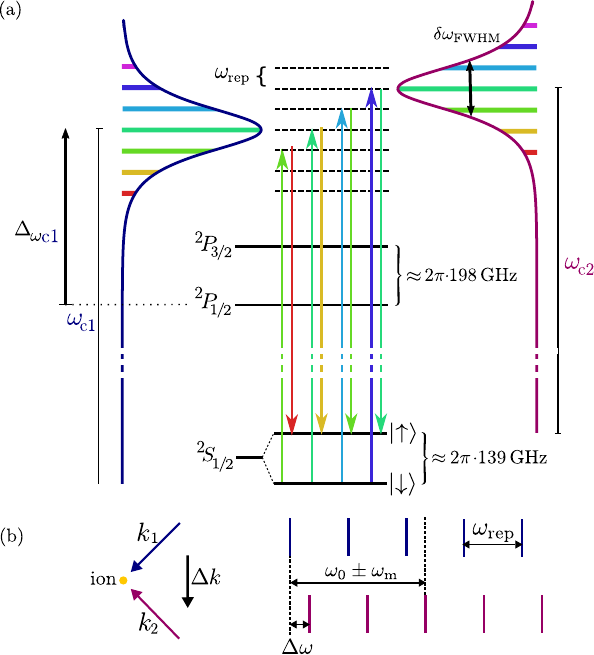}
	\caption{
		Illustration of the two-photon stimulated Raman process implemented by two beams from a single mode-locked laser for \BePlus ions at a ground state sublevel splitting of $\omega_0$($B$\,=\,5\,T)\,$\approx2\pi\cdot139\,$GHz.
		(a) Raman transitions occur by absorption of a photon from off-resonant comb teeth of beam\,1 and stimulated emission into off-resonant comb teeth of beam\,2. We reference detunings to the $^2P_{1/2}$-level, as illustrated for the first comb's central frequency $\omega_{\text{\text{c}1}}$, $\Delta_{\omega{\text{c}1}}$.
		(b) Illustration of the relative frequency shift $\Delta \omega$ between both beams, required to fulfill the Raman resonance condition for internal-motional coupling.
	}
	\label{fig:figure1}
\end{figure}

Laser cooling and manipulation have enabled many groundbreaking experiments in quantum control of atoms and molecules. Prominent applications include precision measurements, frequency metrology and quantum information processing. Whereas continuous wave (CW) lasers are widely established in this context, direct manipulation through mode-locked lasers~\cite{hayes_entanglement_2010} offers unique advantages, enabling a new class of experiments requiring coherent sources spanning large frequency differences. The broad spectrum of regularly spaced, phase-coherent comb modes of pulsed lasers allows to directly bridge large frequency gaps in a controllable, flexible and technically simple way. This scenario is of particular relevance for the control of molecular quantum states~\cite{ding_quantum_2012} as well as for trapped-ion experiments in Penning traps~\cite{brown_geonium_1986, thompson_applications_2009}, where the characteristic high magnetic fields of up to several Tesla cause large Zeeman shifts between atomic states on the order of several hundreds of GHz. Here, the standard approach of stimulated-Raman laser manipulation for quantum control~\cite{wineland_quantum_2003} becomes very difficult to implement using CW lasers. Full control over all quantized degrees of freedom has thus not yet been implemented in Penning traps. The use of pulsed lasers in this context promises to overcome this limitation and enable a new class of quantum logic and quantum information experiments in Penning traps.
\\
Most of the laser coolable atomic ions have optical resonances in the ultraviolet (UV) spectral region, whereas mode-locked lasers~\cite{udem_femtosecond_2009} typically emit in the infrared spectral domain.  Their high intensity, however, enables efficient nonlinear conversion processes, so that a wide range of atomic species can be eventually covered. Due to their light mass and the absence of metastable electronically excited states, \BePlus ions in particular are considered as auxiliary or ``logic'' ions in protocols for sympathetic laser cooling and quantum logic spectroscopy with (anti-)protons in Penning traps~\cite{heinzen_quantum-limited_1990,wineland_experimental_1998} and ultra-precise \textit{g}-factor based tests of fundamental symmetries. Towards this end, here we show numerical simulations of direct frequency comb control of \BePlus ions for a wide range of magnetic field regimes. We introduce novel techniques for the selective and efficient generation of a narrow-bandwidth UV frequency comb with tunable spectral properties and experimentally demonstrate spin-motional control of \BePlus ions using a spectrally optimized optical frequency comb. 
\\
The implementation of quantum control is carried out using two-photon stimulated Raman transitions. The coupling allows full control over internal and motional degrees of freedom and thereby enables sideband (ground state) cooling, quantum information processing and metrology applications of atoms and molecules. In the standard approach, these transitions are driven by two CW laser beams with a frequency detuning equal to the energy splitting between the states to be coupled. The beams are derived either from a single laser source using frequency modulators or from two phase-locked lasers. With increasing energy splitting, this approach becomes intractable due to the lack of efficient high frequency modulators and technical complexities of phase locking two CW lasers with large frequency differences. In contrast, driving the transitions directly using an optical frequency comb allows a coupling across large energy differences without needing a second laser or high frequency modulators, if the comb's spectral bandwidth is broader than the level splitting. In this scenario, the atom undergoes coherent transitions by absorption of a photon from one comb tooth and stimulated emission into another tooth, as illustrated in Fig.~\ref{fig:figure1}(a). Due to the regularity of comb mode spacing, multiple pairs of comb teeth coherently add up and contribute to the process~\cite{hayes_entanglement_2010}. If only internal spin states of the atom are to be coupled (``carrier transitions''), the comb teeth of each pair can be provided by a single laser beam, if the internal-state transition frequency $\omega_0$ is an integer multiple of the laser's repetition rate $\omega_{\text{rep}}$, $\omega_0 = q \cdot \omega_{\text{rep}}$ with $q\in \mathbb{Z}$. If changes in the motional state of the ion are simultaneously pursued (``motional sideband transitions''), the comb teeth of each pair must originate from separate beams, directed onto the ion from different directions and synchronized in time. In this case, the resonance condition for $n$-th order sideband transitions with motional frequency $\omega_m$, $\omega_0 \pm n \cdot \omega_m = \lvert j \cdot \omega_{\text{rep}} \pm \Delta{\omega}\rvert$ with $n,j\in \mathbb{Z}$ is fulfilled by a relative frequency shift between the beam's frequencies $\Delta\omega$, implemented such that teeth from the one beam's spectrum in combination with teeth from the other beam's spectrum can bridge the level splitting, as illustrated in Fig.~\ref{fig:figure1}(b). In contrast to the CW-laser approach, the required frequency shift between the beams does not need to cover the entire level splitting and the optimum shift instead is on the order of only one third of the laser's repetition rate~\cite{mizrahi_quantum_2014}. This is a clear advantage of this type of direct frequency comb control, which previously has been demonstrated with $^{171}\text{Yb}^{+}$ ions for a level splitting near 10\,GHz~\cite{hayes_entanglement_2010} and recently been employed for spectroscopy on carrier transitions in $^{40}\mathrm{Ca}^+$~\cite{solaro_direct_2018}.
\begin{figure}[tb]
	\centering
	\includegraphics[width=1.0\columnwidth]{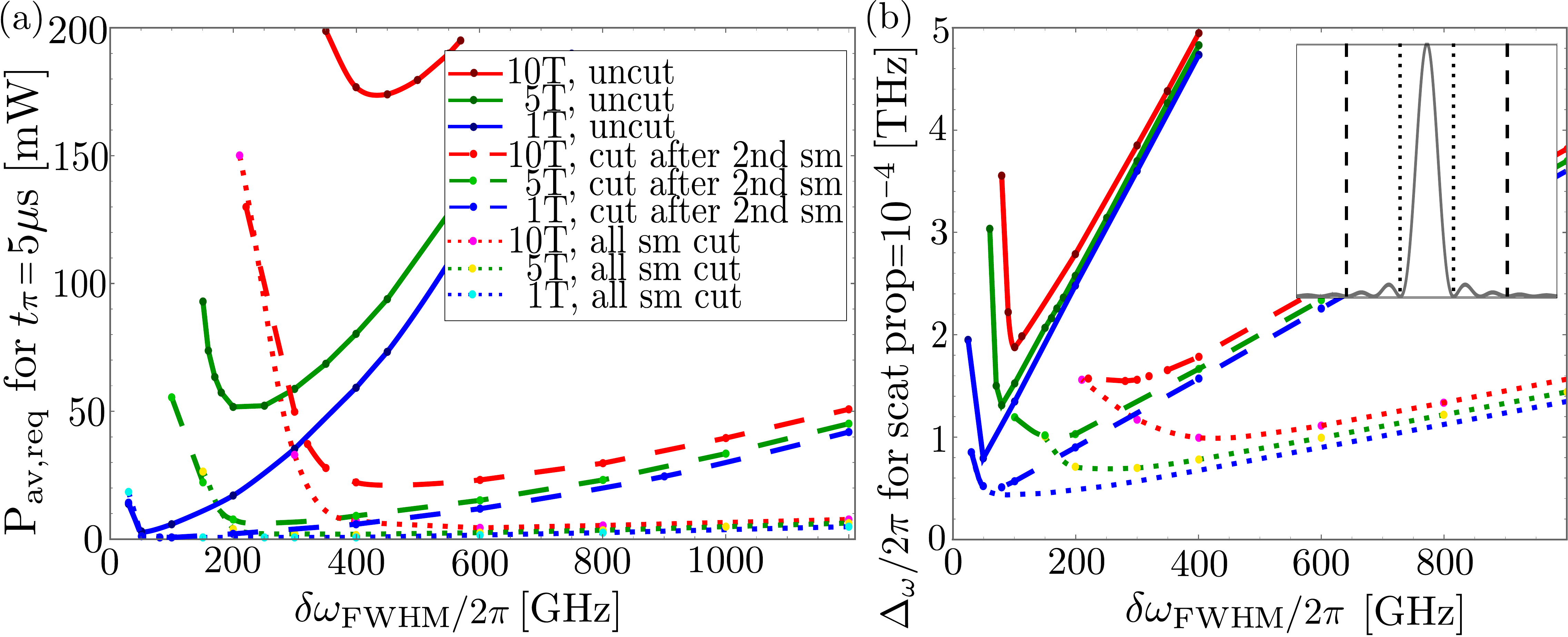}
	\caption{
		Numerical simulations showing the influence of the comb's spectral properties on achievable Raman coupling strengths for different applied magnetic fields.
		(a) Average laser power $P_{\text{av,req}}$ required in order to achieve a spin flip within $t_{\pi}$\,=\,5\,$\mu$s (calculated as in~\cite{hayes_entanglement_2010}) while ensuring the total scattering probability to be $(\Gamma_{\text{total1}} +\Gamma_{\text{total2}})\cdot5\,\mu$s\,=\,$10^{-4}$ as a function of the UV comb's spectral bandwidth $\delta\omega_{\text{FWHM}}$. 
		(b) Corresponding required detuning $\Delta_{\omega}/2\pi$ of the comb's central frequency as a function of $\delta\omega_{\text{FWHM}}$. The total scattering rate of beam $j$ is composed of Raman and (decohering) elastic Rayleigh scattering~\cite{uys_decoherence_2010} according to $\Gamma_{\text{total}_j}$\,=\,$1/2 (\Gamma_{\text{Raman}_j}(\downarrow)+ \Gamma_{\text{Raman}_j}(\uparrow) + \Gamma_{\text{el}_j})$, with $\Gamma_{\text{Raman}_j} $ and $\Gamma_{\text{el}_j}$ calculated following~\cite{uys_decoherence_2010} with the squared electric field amplitudes replaced by the sum over all contributing comb modes. Data for different magnetic fields are shown each for a sinc-shaped spectrum, either unmodified (solid) or spectrally cut outside the main peak (dotted) or the second side-maxima (sm) (dashed). The lines are guides to the eye. Assumed were time-bandwidth limited pulses and a focal radius of 15\,$\mu$m. The sinc-shaped spectrum approximates the unmodified spectrum of our UV frequency comb but is also representative of a more general pulse shape which typically exhibits spectral side maxima.
	}
	\label{fig:figure2}
\end{figure}
\\
In our scenario~\cite{smorra_base_2015,paschke_9be+_2017} we intend to apply the pulsed laser approach for full quantum control of \BePlus ions in the environment of a Penning trap at a magnetic field of 5\,T leading to a Zeeman level splitting near 140\,GHz. A particular challenge for the implementation arises from beryllium's atomic structure. The atomic levels for implementing quantum control are chosen within the sublevels of beryllium's single ground state $S$-level, $1s^22s$, whereas the stimulated-Raman coupling is facilitated by off-resonant coupling to beryllium's electronically excited $P$-state, $1s^22p$. Here, the small excited state fine-structure splitting of only 198\,GHz between the $^2P_{1/2}$ and $^2P_{3/2}$ levels requires a Raman detuning outside the $P$-level manifold in order to avoid resonant excitation of the $P$-levels, which would be followed by spontaneous emission decay. For this atomic configuration the comb's spectral properties, such as the spectral envelope shape and its width, need to be precisely controlled in order to achieve high Raman coupling strengths, while simultaneously guaranteeing reasonably low scattering rates. 
\\
Figure~\ref{fig:figure2} illustrates the influence of the comb's spectral properties on achievable Raman coupling strengths for representative magnetic fields $B$ from 1\,T to 10\,T with corresponding transition frequencies $\omega_0(B)$ from $2\pi \cdot$28\,GHz to  $2\pi \cdot$280\,GHz. For each magnetic field the comb's spectral properties in general set the minimum Raman detuning required in order to control the scattering rate, which strongly influences and limits the maximum achievable Raman coupling strength. For an unmodified spectrum (solid lines) the optimal spectral bandwidth is only slightly broader than the respective level splitting at a given magnetic field, as demonstrated in Fig.~\ref{fig:figure2}(a). For narrower spectra too few pairs of comb teeth exist that contribute to the Raman process, while for broader spectra a strong loss of achievable Raman coupling occurs because the detuning required to suppress spontaneous scattering below a certain threshold significantly increases, as shown in Fig.~\ref{fig:figure2}(b). This loss of achievable coupling strength due to the increasing required detuning strongly depends on the comb's specific spectral envelope. In particular, any power contained in spectral wings close to resonance strongly enhances decoherence through spontaneous emission without equally contributing to the coupling strength. A significant improvement of the achievable coupling strength (for a given target scattering rate) can hence be implemented through spectral pulse shaping by blocking the inefficient spectral tails of the UV pulse close to resonance. The absence of these outer-lying frequency components allows for much smaller detunings, which has a strong impact on maximizing achievable Raman coupling strengths, as shown in Fig.~\ref{fig:figure2} for exemplary modified spectra (dashed and dotted lines). In addition the comb's spectral bandwidth can be chosen in a much wider range than for unmodified spectra and achievable couplings become almost identical over a wide range of magnetic fields; both properties strongly improve the flexibility of this approach.
\\
In order to obtain an efficient Raman coupling for \BePlus ions it is thus mandatory to combine a narrow-bandwidth frequency comb generation with a technique for spectral control allowing to block the comb's spectral wings. The resulting comb then provides enormous flexibility enabling efficient quantum control of \BePlus ions at nearly any experimentally relevant conditions.
\\
For the experimental implementation of efficient and versatile quantum control we have developed a narrow-bandwidth UV frequency comb with tunable spectral properties (details in~\cite{paschke_9be+_2017}). The system is based on a custom-built femtosecond frequency comb operating near 626\,nm with a tunable repetition rate near 100\,MHz. The output is then frequency doubled to the UV in order to generate the desired wavelength for beryllium's optical resonance near 313\,nm. Control of the spectral properties is realized by the combination of a nonlinear spectral compression technique implemented during second harmonic generation (SHG) into the UV and a subsequent blocking of the comb's spectral wings. 
\begin{figure}[tb]
	\centering
	\includegraphics[width=0.99\columnwidth]{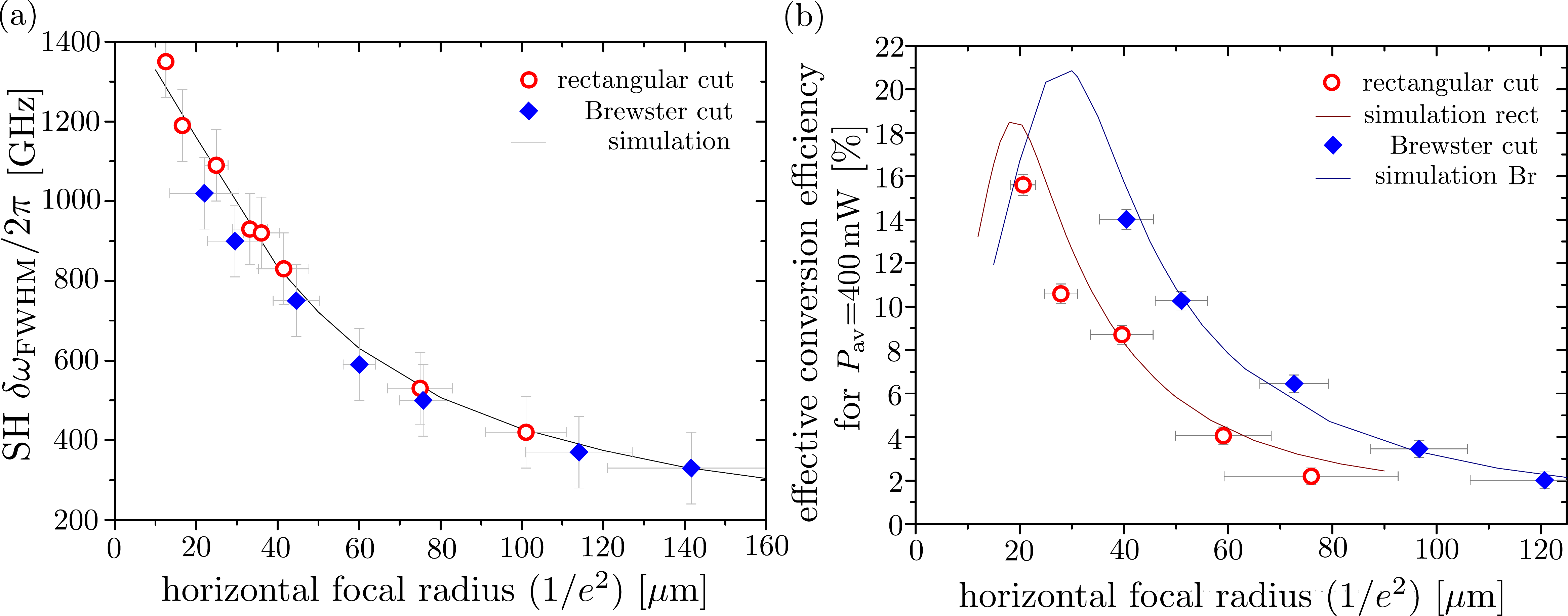}
	\caption{
		Focus-dependent spectral compression technique implemented during SHG from broadband pulses near 626\,nm into narrow-bandwidth pulses near 313\,nm using a 4.5\,mm long BiBO crystal in single pass configuration.
		(a) Resulting UV SH bandwidth as a function of the focal radius of the FF beam along the relevant direction of spatial walkoff (horizontal).
		(b) Effective conversion efficiencies for an average input power of 400\,mW as a function of the horizontal focal radius of the FF beam. For the rectangular cut crystal a circular focus was present, whereas for the Brewster cut crystal the focal dimension in transverse direction by geometry was reduced by approximately a factor of $n_{\text{BiBO}}\approx 1.88$, thereby enhancing conversion efficiencies for a given horizontal radius. The solid lines correspond to simulation results obtained with the model described in~\cite{lang_impact_2013}. We assume transform-limited gaussian input pulses with an initial bandwidth of $\Delta \lambda=1.5$\,nm in (a) and $\Delta \lambda=0.9$\,nm in (b).
	}
	\label{fig:figure3}
\end{figure}
\begin{figure}[tb]
	\centering
	\includegraphics[width=0.95\columnwidth]{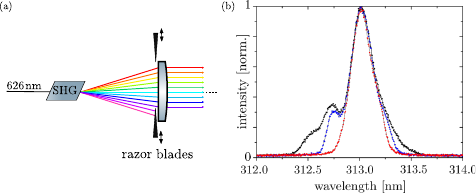}
	\caption{
		(a) Illustration of spectral pulse modification technique. Razor blades located directly behind the SHG crystal address and block the comb's unwanted spectral wings, which are spatially separated due to spatio-temporal coupling effects causing an angularly dispersed UV signal. 
		(b) Sample modified UV spectrum. In black the unfiltered UV spectrum is shown, with its asymmetric shape dominated by the spectral shape of the customized fiber-based frequency comb.
		The blue and red spectra correspond to different positions of the razor blades independently moved (with decreasing distance) on each side of the spatial walkoff plane without affecting the intensity of the main peak.
	}
	\label{fig:figure4}
\end{figure}
\\
The nonlinear spectral compression technique applied to our system (cf. supplementar material) uses the SHG process to transfer the energy of broadband fundamental frequency (FF) femtosecond pulses into narrow-bandwidth second harmonic (SH) ones. Due to different group velocities of the FF and the SH pulses inside the crystal, the SH pulse is temporally broadened during propagation, which is directly accompanied by a spectral narrowing. In the spectral domain this process is related to the limited phase-matching bandwidth of the nonlinear crystal in terms of the SHG process expressed as $\Delta \omega_{\text{SH}} \approx 2 \pi \cdot\frac{0.886}{\lvert GVM \rvert \cdot L}$~\cite{marangoni_narrow-bandwidth_2007}, with $GVM$ denoting the group velocity mismatch between the FF and the SH pulse and $L$ the crystal length. In case the pulses remain spatially overlapped during propagation within the nonlinear crystal, the resulting SH bandwidth can thus be controlled by the crystal length, as shown in previous work for the generation of tunable narrow-bandwidth pulses in the visible spectral region~\cite{marangoni_narrow-bandwidth_2007}. Here we include the effect of spatial walkoff, where the FF and SH pulses travel in different directions while propagating through the crystal. We have theoretically and experimentally investigated spatio-temporal coupling effects onto the nonlinear conversion and spectral compression process and show the UV pulse output bandwidth for a given nonlinear crystal length to be selectively tunable by varying the focusing conditions of the FF pulse. For implementation, the nonlinear crystal bismuth triborate~\cite{hellwig_exceptional_1998,hellwig_linear_2000,ghotbi_optical_2004}, BiBO, is chosen as it combines a large $GVM$ (984\,fs/mm~\cite{smith_snlo_2013}) for efficient spectral compression with a high nonlinearity ($d_{\text{eff}}=3.36$\,pm/V~\cite{smith_snlo_2013}) for high conversion efficiencies. Results for an exemplary crystal length of 4.5\,mm are presented in Fig.~\ref{fig:figure3}. The spectral properties of the SH pulse are shown to strongly depend on the focal beam dimension along the direction of spatial walkoff and can be tuned over a wide range of bandwidths for a given crystal length. With increasing focal radii the effective crystal length over which the fundamental and the SH beams are overlapped increases and thus a stronger spectral compression is achieved, while the peak intensities and the corresponding conversion efficiencies drop. A selective and efficient generation of narrow-bandwidth UV pulses with a wide range of bandwidths is thus achieved by choosing appropriate combinations of crystal lengths and focusing conditions.
\\
For the subsequent spectral pulse modification, we utilize the angular spectral chirp resulting from spatio-temporal couplings occurring during the SHG process. Here, different spectral components of the generated UV signal leave the crystal under different angles. Rather than relying on a pulse shaping arrangement involving UV gratings with high transmission losses, this allows us to block the comb's undesired spectral wings in a straightforward way by using razor blades located closely behind the crystal, as shown in Fig.~\ref{fig:figure4}. 
\\
The experimental demonstration of stimulated-Raman laser control of \BePlus ions using the spectrally modified comb system has been carried out in a surface-electrode experiment at an externally applied magnetic field of 22.3\,mT~\cite{wahnschaffe_single-ion_2017,paschke_9be+_2017} because our Penning trap apparatus with a 5\,T field is still in commissioning. Furthermore, for $\delta\omega_\mathrm{FWHM}/2\pi>300\,\mathrm{GHz}$, achievable coupling strengths are similar for 22.3\,mT and for 5\,T as shown in Fig.~\ref{fig:figure5} (a), demonstrating the flexibility of our approach. In the 22.3\,mT case, in principle even lower power may be required if the spectrum can be made narrower, down to the qubit splitting of $\approx1\,\mathrm{GHz}$; in that case, however, the coupling strength for the 5\,T case would be significantly reduced. Furthermore, the required length of the nonlinear crystal to achieve such extremely small bandwidths would be unrealistic (tens of centimeters). 
\begin{figure}[tb]
	\centering
	\includegraphics[width=0.95\columnwidth]{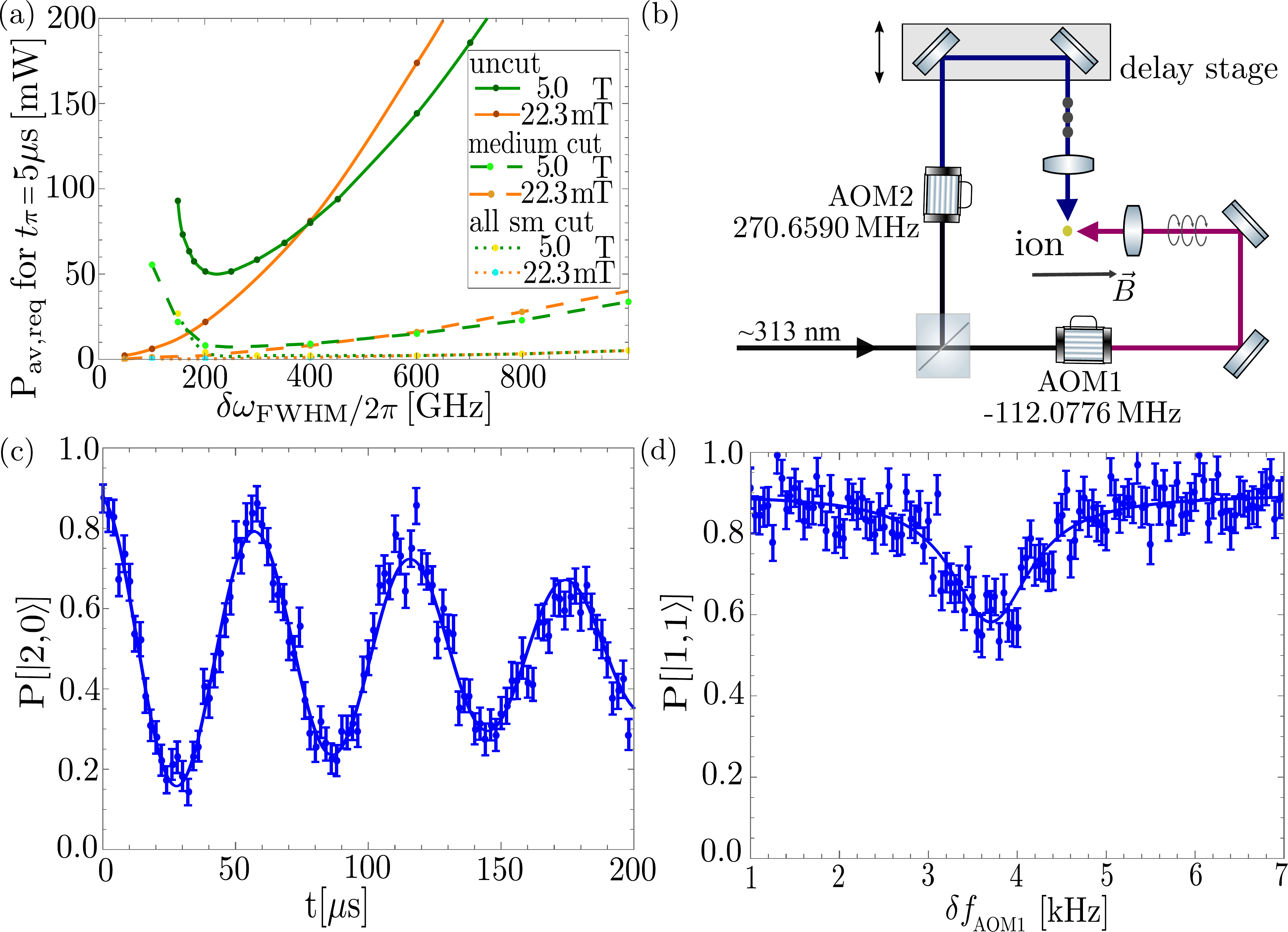}
	\caption{
		Direct pulsed laser quantum control of \BePlus ions.
		(a) Numerical simulations (similar to Fig.~\ref{fig:figure2}(a)) comparing achievable Raman coupling strengths for different magnetic field regimes. In green for the Paschen-Back regime with a magnetic field of 5\,T and in orange for the low-field Zeeman regime with a magnetic field of 22.3\,mT, each with differently modified spectra as discussed in Fig.~\ref{fig:figure2}(a).
		(b) A spectrally tailored UV pulse train is divided into two beam paths and focused onto the ion from orthogonal directions and synchronized in arrival time using a temporal delay stage. 
		(c) Single beam induced carrier Rabi oscillations on the $|2,0\rangle\leftrightarrow |1,0\rangle$ transition implemented using only the first Raman beam with AOM1. 
		Shown is the population of the $\lvert2,0\rangle$ state as a function of the laser probe duration. Decoherence is expected to be dominated by the strong magnetic field noise sensitivity of the transition of 12.53\,MHz/mT. The comb's applied spectral bandwidth of the main peak was	$\delta\omega_{\text{FWHM}}\approx 2\pi \cdot 940\,$GHz and its detuning $\Delta_{\omega{\text{c}1}} \approx 2\pi \cdot 1.26\,$THz, with both values optimized according to the achievable UV pulse generation, shaping and Raman coupling performance. 
		(d) Resonance scan of a red sideband transition on the $\lvert2,1\rangle\leftrightarrow \lvert1,1\rangle$ transition for the ion's axial motional mode ($\omega_{\text{m,axial}} =2\pi \cdot 0.892$\,MHz), driven by the interaction of both Raman beams with a frequency shift of $\Delta\omega_{\text{AOM}}= 2\pi \cdot 382.7366\,$MHz. Shown is the population of the $\lvert1,1\rangle$ state as a function of a frequency shift applied to AOM1. The red sideband transition probability is due to the Doppler cooled motional state of the ion.
	}
	\label{fig:figure5}
\end{figure}
\\
Internal-state control has been implemented on beryllium's ground state sublevel transition between the $|F=2,m_F=0\rangle$ and the $\lvert F=1,m_F=0\rangle$ state. A single train of spectrally tailored pulses was focused onto the ion, switched on and off using a single-pass acousto-optic modulator (AOM). Fulfilling the resonance condition $\omega_0=2 \pi \cdot 1397.8372\,\text{MHz} = 14 \cdot \omega_{\text{rep}} $, the interaction with the pulsed Raman beam causes the atom to undergo a coherent evolution resulting in Rabi oscillations, as shown in Fig.~\ref{fig:figure5}(c). The achieved $\pi$-time of $t_{\pi}=29.04\,\mu \text{s} \pm 0.07\,\mu \text{s}$ coincides with simulations under given experimental conditions~\cite{paschke_9be+_2017}. The benefit of the pulse shaping is evident when, after choosing laser detuning and bandwidth, by moving in the razor blades, the contrast of Rabi oscillations improves until limited by ambient magnetic field fluctuations (cf. caption and supplementary information) without apparent loss of oscillation speed. This is also evident from shining in the beam such that the two-photon resonance condition is not met. Without pulse shaping, spontaneous scattering leads to a strong depletion of the initial state (cf. supplementary information).
\\
Internal-motional coupling has been demonstrated on the first-order magnetic field-insensitive $|F=2,m_F=1 \rangle\leftrightarrow\lvert F=1,m_F=1\rangle$ transition. The output of the spectrally modified UV comb was divided in two beams which were focused onto the ion from orthogonal directions and synchronized in arrival time, as shown in Fig.~\ref{fig:figure5}(b). By sending each beam through a single-pass AOM with adjusted frequency, the resonance condition for a red sideband transition on the axial mode, $\omega_0-\omega_m=2 \pi \cdot 1081.6551\,\text{MHz} = 7 \cdot \omega_{\text{rep}} + \Delta\omega_{\text{AOM}} $ was fulfilled, as shown in Fig.~\ref{fig:figure5}(d). Note that the spectrum was obtained for a Doppler cooled ion, and therefore we do not observe full contrast on the sideband transition.

In summary we have demonstrated internal state control and internal-motional state coupling of \BePlus ions implemented by stimulated-Raman control using an optical frequency comb. This represents a key enabling step for full quantum control of \BePlus ions in quantum information and quantum logic inspired precision experiments in Penning traps. Towards this end, we have developed novel techniques for the generation of narrow-bandwidth UV pulses with tunable spectral properties fulfilling all requirements for efficient and flexible quantum control. These techniques
might further find use in similar scenarios of interest, such as cooling and manipulation of ultracold molecules or time-resolved spectroscopy.

\begin{acknowledgments}
We acknowledge funding from QUEST, PTB, LUH, Laserlab Europe, ERC StG ``QLEDS'' and DFG through SFB 1227 ``DQ-mat''. 
\end{acknowledgments} 
\bibliography{qc}
\end{document}


\bibliographystyle{apsrev}

\section{Supplementary Material} 

\subsection{Spectral compression without spatial walkoff}

\begin{figure} [t]
	\centering{
		\includegraphics[width=1.0\columnwidth]{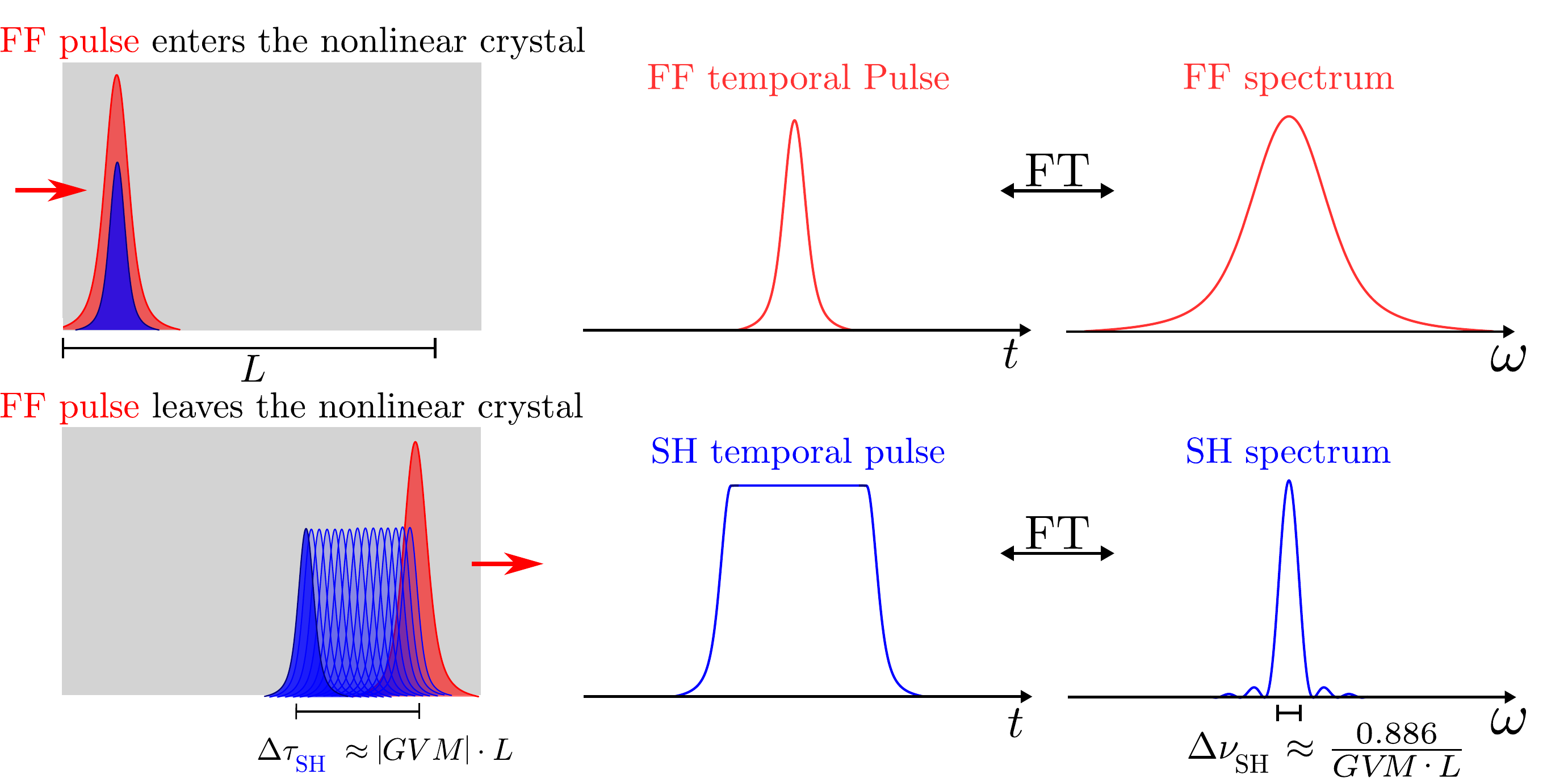}
	}
	\caption{Illustration of the spectral compression process during second-harmonic generation without spatial walkoff. If phase-matching conditions are fulfilled, the fundamental frequency pulse (FF), shown in red, continuously generates a second-harmonic wave (SH), shown in blue, while travelling through the nonlinear crystal. Due to the smaller group velocity, the second-harmonic signal lags behind in time about $\delta t= GVM \cdot L$ at the moment when the fundamental pulse exits the nonlinear crystal. This temporal separation corresponds to the temporal broadening of the generated SH signal, $\Delta \tau_{\text{SH}}$, which is directly accompanied by a spectral compression, as shown on the right. 
	}
	\label{fig:principleGVM}
\end{figure}

The spectral compression technique during the second-harmonic generation, without appearance of spatial walkoff, can be easily understood in the time-domain picture. The spectral compression directly results from the temporal broadening of the generated signal, occurring due to the group-velocity mismatch and consequent temporal walkoff between the fundamental pulse (FF) and the second-harmonic pulse (SH) inside the nonlinear crystal. When the fundamental pulse enters the crystal it will generate a SH pulse, if phase-matching conditions for SHG are fulfilled. At this moment the FF and SH pulses overlap in time. When the FF pulse has traveled through the crystal and exits the end facet, due to its smaller group velocity, the initially generated SH pulse lags behind in time by
\begin{equation}
\delta t\approx \lvert GVM \rvert \cdot L\,,
\label{eq:tempbw}
\end{equation}
with the crystal length $L$ and the group velocity mismatch $GVM=1/v_\mathrm{GFF}-1/v_\mathrm{GSH}$, where $v_\mathrm{GFF}$ and $v_\mathrm{GSH}$ are the group velocities of the FF and SH pulses, respectively. As the FF pulse continuously generates a SH wave while traveling through the crystal, this temporal separation approximately corresponds to the temporal width of the resulting SH signal, as depicted in Fig.~\ref{fig:principleGVM}:
\begin{equation}
\delta t\approx \Delta t_{\text{SH}} 
\end{equation}
The Fourier transform of this broad temporal pulse directly yields the narrow spectrum. The spectral compression is hence directly linked to the amount of temporal broadening of the generated SH output signal. Without the appearance of spatial walkoff the resulting SH temporal pulse profile is approximated to be rectangularly shaped due to the one-dimensional temporal broadening as illustrated in Fig.~\ref{fig:principleGVM}. The resulting width of the corresponding spectrum is therefore given by~\cite{fejer_quasi-phase-matched_1992,marangoni_narrow-bandwidth_2007}:
\begin{equation}
\Delta \nu_{\text{SH}} \approx \dfrac{0.886}{\lvert GVM \rvert \cdot L}
\label{eq:specbw}
\end{equation}
The spectral bandwidth of the generated SH pulse hence depends on the GVM and the crystal length $L$. The longer the crystal and the higher the GVM, the narrower the output spectrum. The approximation of a rectangular temporal profile also determines the SH spectral shape, which in this case is given by a sinc-shaped spectrum. As a consequence of this limited acceptance range, frequency doubling of broad spectra is typically performed using short crystals featuring a low GVM. In the present work, on purpose, long crystals with large GVM are chosen, such that the acceptance range, and hence the resulting SH output bandwidth, equals the desired narrow SH bandwidth.

In terms of the SHG process therefore only a small fraction of the FF pulse is used and converted into the SH one for the generation of narrow spectra. Based on this, a very low conversion efficiency would be expected. However, to first order also phase matching for intrapulse sum-frequency generation (SFG) is fulfilled, generating the same frequencies as resulting from the frequency doubling. 

For the SHG process, the second-harmonic frequency, $\omega_{\text{SH}} = 2 \omega_{\text{pm}}$, with pm denoting the phasematched fundamental center frequency, is generated by:
\begin{equation}
\omega_{\text{pm}} + \omega_{\text{pm}} = 2 \omega_{\text{pm}},
\end{equation}
with the fundamental frequency $\omega_{\text{FF}}= \omega_{\text{pm}}$. The corresponding phase-matching condition for the wave vectors
\begin{equation}
k(\omega_{\text{pm}}) + k(\omega_{\text{pm}}) = k(2 \omega_{\text{pm}})
\end{equation}
is assumed to be fulfilled. The intrapulse SFG process generates the same frequency $\omega_{\text{iSF}} = 2 \omega_{\text{pm}}=\omega_{\text{SH}}$ by mixing spectral components of the FF spectrum which are symmetric around the phase-matching frequency $\omega_{\text{pm}}$ about $\pm \Delta \omega$: 
\begin{equation}
(\omega_{\text{pm}} + \Delta \omega) +  (\omega_{\text{pm}} - \Delta \omega) = 2 \omega_{\text{pm}}
\end{equation}
The SFG phase-matching condition
\begin{equation}
k(\omega_{\text{pm}} + \Delta \omega)+k(\omega_{\text{pm}} - \Delta \omega) = k(2 \omega_{\text{pm}} )
\end{equation}
is approximately simultaneously fulfilled. 

The corresponding phase-matching acceptance bandwidth is found to be~\cite{marangoni_narrow-bandwidth_2007}
\begin{equation}
\Delta \nu_{\text{SFG}} \approx \sqrt{\dfrac{0.886}{2 \pi L k''_{\text{FF}}} }
\label{eq:specbwipsfg}
\end{equation}
with $k''_{\text{FF}}$ denoting the group velocity dispersion of the fundamental frequency. This acceptance range of intrapulse SFG is typically broader than the acceptance range of the simultaneously occurring SHG process, whereby this spectral compression technique allows to efficiently convert broadband FF pulses into frequency doubled narrow-band pulses. The width of the resulting pulses is, however, determined by the acceptance range of the SHG process.

This spectral compression method has been successfully demonstrated for the generation of narrow-bandwidth pulses in the visible region~\cite{marangoni_narrow-bandwidth_2007}. A high conversion efficiency has been achieved with long periodically poled crystals, which are effectively free of spatial walkoff and therefore guarantee an overlap of the SH and the FF pulses over the entire crystal length. For the generation of UV light near 313\,nm no practical solutions exist to directly avoid spatial walkoff. The spatial walkoff and the resulting spatio-temporal coupling effects strongly influence the compression process and have to be taken into account in order to selectively and efficiently generate narrow-bandwidth UV pulses.

\subsection{Spectral compression with spatial walkoff}

In the presence of spatial walkoff, the FF and the SH pulses not only lose overlap due to the temporal walkoff between them because of their different group velocities, they are further separated in space due to their different propagation directions. This strongly affects the resulting SH pulse profile during the SHG and hence also the spectral compression process. In general less temporal broadening and therefore less spectral compression can be achieved for the same crystal length, compared to the case without spatial walkoff. Furthermore the resulting SH temporal shape is distorted and cannot be approximated by a rectangular pulse shape anymore. Due to the different propagation directions, combined with the different group velocities, the resulting SH profile experiences a pulse front tilt~\cite{akturk_general_2005}. 

\begin{figure} [t]
	\centering{
		\includegraphics[width=\columnwidth]{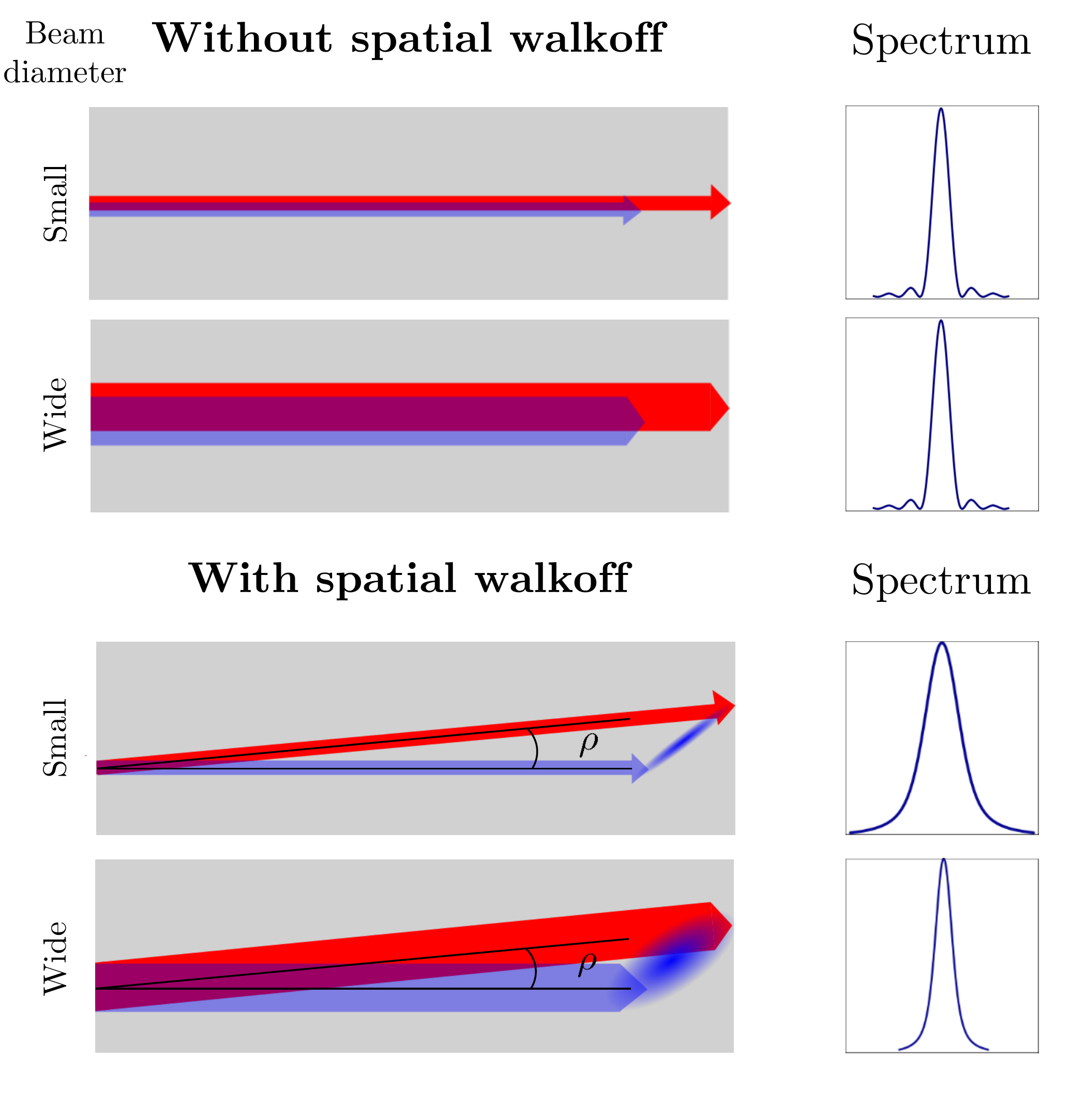}
	}
	\caption{Origin of the pulse front tilt of the second harmonic (SH) signal, caused by simultaneous spatial and temporal walkoff during the SHG. The initial FF propagation direction and hence the direction of both wavevectors of the FF and SH inside the crystal is left to right. The spatial walk-off takes place in the perpendicular orientation of the nonlinear crystal (up-down). The fundamental beam direction is shown in red and experiences a spatial walkoff about the angle $\rho$ compared to the propagation direction of the SH beam, shown in blue. The combination of different propagation directions and different group velocities between the FF and the generated SH signal leads to a tilted pulse front of the SH signal when leaving the nonlinear crystal, indicated by the blue ellipse.}
	\label{fig:walkoff}
\end{figure}

The generation of the temporal SH profile including the influence of spatial walkoff can be understood in the time-domain picture, analogous to the simple picture presented in Figure \ref{fig:principleGVM}. When the FF pulse enters the crystal it will generate a SH pulse, if phase-matching conditions for SHG are fulfilled. At this moment the FF and SH pulses overlap in time and in space. When the FF pulse has traveled through the crystal along its direction of propagation and exits the crystal end facet, the initially generated SH pulse lags behind in time due to its smaller group velocity. Due to the different propagation direction compared to the FF, the initially generated SH pulse is not only located at a different position along the direction of propagation,
but it is also located at a different transverse position inside the nonlinear crystal. As the FF pulse continuously generates SH waves while traveling through the crystal, the combination of temporal and spatial walkoff causes a tilted pulse front of the generated SH signal when leaving the nonlinear crystal.

The precise temporal profile not only depends on the walkoff angle $\rho$, but also on the GVM and the propagation distance of the FF pulse inside the nonlinear crystal. Furthermore the focusing strength is of great importance, influencing the temporal profile by affecting the effective interaction length between the FF and SH signal. Whereas in the case without spatial walkoff the interaction length is independent of the focusing strength, leading to the same spectral results for any focusing conditions, the interaction length is in general reduced by the effect of spatial walkoff. The limitation of the interaction length becomes the stronger, the tighter the focusing is. For weak focusing, corresponding to large beam sizes, the interaction length is larger, whereas in the limit of very large beam sizes an interaction over the entire crystal length can possibly be achieved, approximately leading to similar results as in the case without spatial walkoff. This corresponds to the large focal radius limit of Fig.~3(a) in the main text. 

\subsection{Coherence of carrier Rabi oscillations}

\begin{figure}[t]
	\includegraphics[width=0.9\columnwidth]{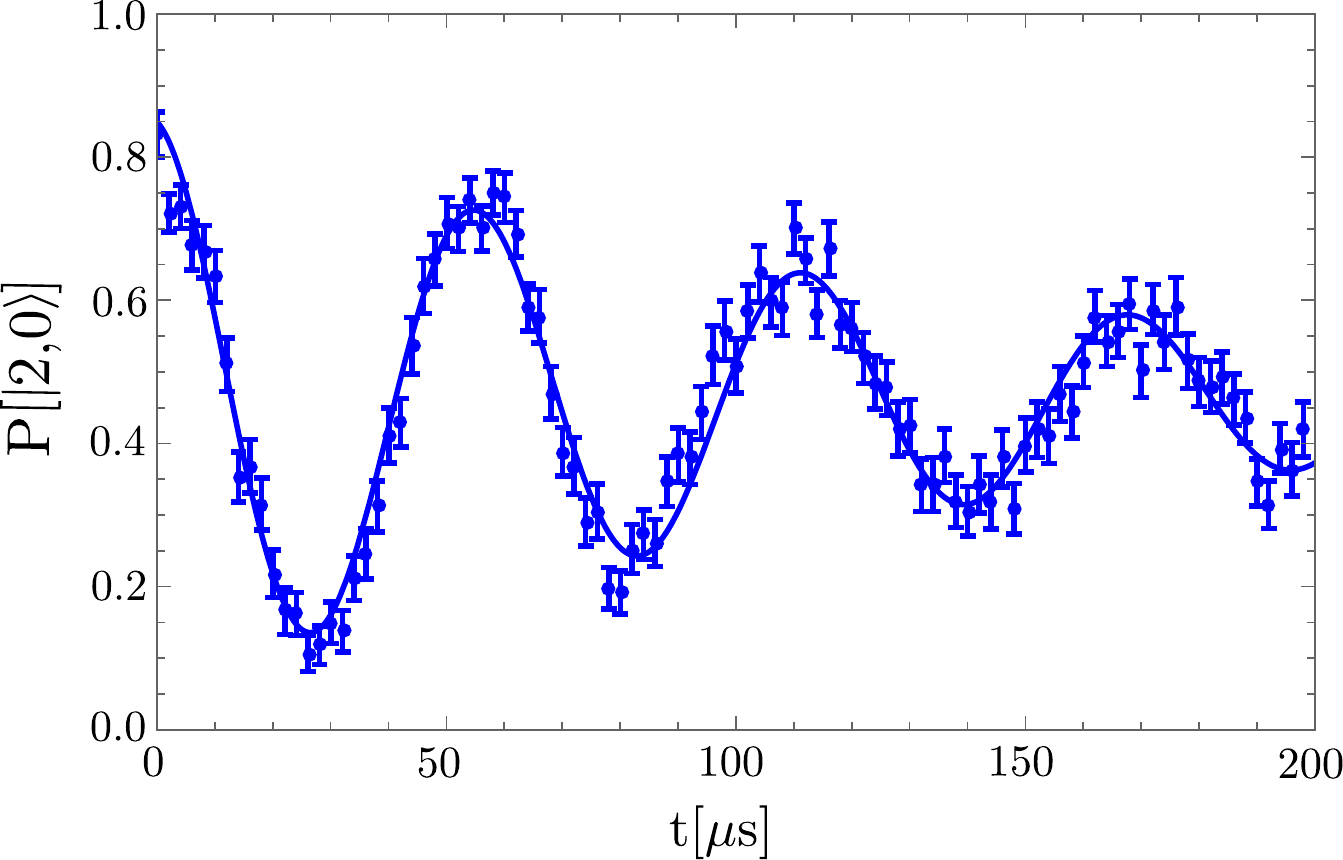}
	\caption{Carrier Rabi oscillations using a microwave field, to be compared to Fig.~5(c) of the main text, and showing the same level of decoherence, induced by slow ambient magnetic field fluctuations. }
	\label{fig:slowMWflop}
\end{figure}

To demonstrate the origin of the decoherence evident in the carrier Rabi oscillations in Fig.~5(c) of the main text, in Fig.~\ref{fig:slowMWflop}, we show carrier oscillations by a microwave magnetic oscillatory field resonant with the same transition and where the Rabi rate was chosen to match the one from the pulsed laser system. The plot in the main text, once fitted with a dampened sinusoid, has a decoherence time of $\tau_{1}=224 \pm 25 \, \mu \text{s}$ while the microwave driven oscillation shown here has decoherence time  $\tau_{2}=141 \pm 11 \,\mu \text{s}$. Since the coherence time for the comb driven oscillations is higher, possibly due to day to day change in experimental conditions, this indicate that any source of decoherence affecting it is lower than the ones present in the microwave driven oscillations where photon scattering from the frequency comb is not a possible. Both plots indicate roughly the same level of decoherence, showing that the source is unrelated to the drive mechanism and most likely caused by ambient magnetic field fluctuations and the use of a magnetic-field sensitive transition.\\

\subsection{Spontaneous scattering}

\begin{figure}[h]
	\includegraphics[width=0.9\columnwidth]{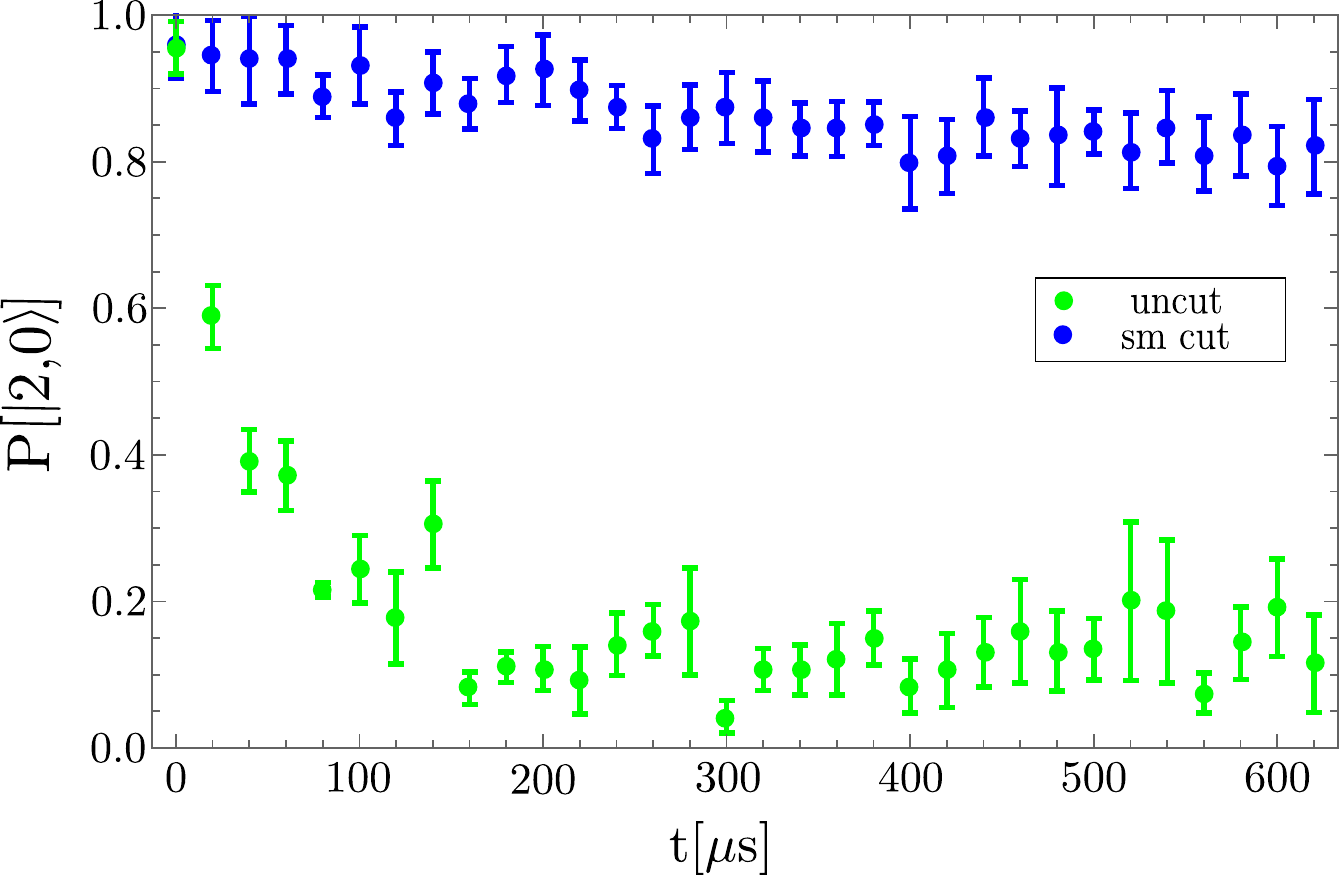}
	\caption{Population depletion of the  $|F=2,m_F=0\rangle$ state due to spontaneous scattering as a function of the time for which the beam illuminates the ion. In green for an unmodified spectrum, in blue after spectral pulse shaping. The beam is off-resonant from the carrier transition to detect only the population depletion, but not Rabi oscillations.}
	\label{fig:cut_uncut_comparison}
\end{figure}

In the case of $^9$Be$^+$ at 22.3\,mT the modification of the frequency comb spectral shape is a fundamental requirement to suppress spontaneus scattering. The actual spectral shape used in the experiments is shown in Fig.~4(b) of the main text, where a broad second peak at lower frequencies is clearly visible in the unmodified spectrum. These components cause a strong spontaneous scattering leading to an effective loss of population from the  $|F=2,m_F=0\rangle$ state.  In Fig.~\ref{fig:cut_uncut_comparison} one can see that if the spectral shape is left unmodified most of the population is lost after ca. $100\,\mu \text{s}$ that both Raman beams are turned on off-resonantly with respect to the carrier transition. Population is redistributed by spontaneous scattering among the $S_{1/2}$ ground state sublevels. The data was taken with a lower power compared to the data in the main text. It is therefore clear that an uncut spectrum would lead to an even faster loss of population which would dominate over Rabi oscillations.

\bibliography{../qc}